\documentclass[preprintnumbers,superscriptaddress,nofootinbib]{revtex4}
\usepackage{amssymb}
\usepackage{amsmath}
\usepackage[dvips]{graphicx,psfrag}

\renewcommand{\thefootnote}{\#\arabic{footnote}}
\newcommand{\slashp}{\not{\hbox{\kern-3pt $P$}}}
\newcommand{\slashs}{\not{\hbox{\kern-3pt $S$}}}

\begin{document}

\renewcommand{\thefootnote}{\fnsymbol{footnote}}
\setcounter{footnote}{0}

\def\thefootnote{\fnsymbol{footnote}}

\preprint{hep-ph/0210448}

\title{Single Heavy MSSM Higgs Production at  $e^+e^-$ Linear Collider
}

\author{Shufang Su}
\email{shufang@theory.caltech.edu}
\affiliation{California Institute of Technology, Pasadena, 
California 91125, USA}

\begin{abstract}
We briefly review the single heavy Higgs production at high energy 
$e^+e^-$ linear collider, $\gamma\gamma$ collider and $e^-\gamma$ collider.
We present the recent results for $e^+e^-\rightarrow W^{\pm}H^{\mp}$ 
in the Minimal Supersymmetric Standard Model and preliminary results
for $e^+e^-\rightarrow \nu \bar\nu A^0$.
\end{abstract}

\maketitle


The measurement of the properties of 
the heavy Higgs bosons  $H^0$, $A^0$ and $H^{\pm}$ of 
Minimal Supersymmetric Standard Model (MSSM)
could tell us a lot about the MSSM parameters.  The mass of the heavy Higgses 
are related to the overall scale $m_{A^0}$ (CP-odd Higgs mass), with
a distinctive mass relation between 
$m_{A^0}$, $m_{H^0}$ and $m_{H^{\pm}}$ in MSSM. 
The heavy Higgs couplings to the fermions are sensitive to 
$\tan\beta$ (the ratio of the two Higgs vacuum expectation values)
and Yukawa couplings, and its couplings to the sfermions are 
sensitive to the trilinear $A$-terms.  Unlike the light CP-even Higgs 
$h^0$, whose couplings become increasingly insensitive to the MSSM parameters
in the decoupling limit ($m_{A^0} \gg m_Z$), 
the couplings of the heavy Higgs bosons 
are always sensitive to the MSSM parameters.
Thus, measurements of the properties of the heavy MSSM Higgs bosons 
are very valuable, especially in the decoupling limit.

Discovery of the heavy Higgs bosons, however, 
poses a special challenge at future colliders, contrary
to the case of the light CP-even Higgs $h^0$, when the discovery 
is almost certain at current and future collider experiments.
Run~II of the Fermilab Tevatron, now in progress, has a limited 
reach for the neutral heavy MSSM Higgs bosons $H^0$ and $A^0$
at small $m_{A^0}$ and large $\tan\beta$.  
The charged Higgs boson $H^{\pm}$ can be discovered in top quark decays
for $m_{H^{\pm}} \lesssim m_t$ and large $\tan\beta$ 
\cite{HiggsatRunII,GuchaitHpm}.
The CERN Large Hadron Collider (LHC) has a much greater reach for heavy
MSSM Higgs boson discovery at moderate to large values of $\tan\beta$
via the process of $H^0$ and $A^0$ decays 
to $\tau$ pairs \cite{AtlasTDR,CMS} and 
the charged MSSM Higgs boson $H^{\pm}$ mode:  
$gb \to t H^-$ with $H^- \to \tau \bar\nu$~\cite{AtlasH+,CMS}.
The absence of a Higgs boson discovery at the CERN LEP-2 experiments
implies that $0.5 < \tan\beta < 2.4$ and $m_{A^0} < 91.9$ GeV
are excluded at 95\% confidence level \cite{LEP2}.
This leaves a wedge-shaped region of parameter space at moderate $\tan\beta$ 
in which the heavy MSSM Higgs bosons could be missed at the 
LHC.

At a future high energy $e^+e^-$ linear collider (LC), the heavy Higgs
bosons will be produced in pairs, if it is kinematically allowed.
The dominant production modes are $e^+e^- \to H^0 A^0$ and 
$e^+e^- \to H^+H^-$.
At large $m_{A^0}$, $m_{A^0} \simeq m_{H^0} \simeq m_{H^{\pm}}$
up to mass splittings of order $m_Z^2/m_{A^0}$, so that the pair-production
modes are kinematically allowed only if $m_{A^0} \lesssim 0.5 \sqrt{s}$.
In particular, the pair-production modes are limited to 
$m_{A^0} \lesssim 250$ GeV ($m_{A^0} \lesssim 500$ GeV) at a LC with
$\sqrt{s} = 500$ GeV ($\sqrt{s} = 1000$ GeV).

In this talk we consider the production of one of the heavy Higgs bosons 
in association with lighter SM particles at the LC.
While the cross sections for such production modes are typically very
small, they offer the possibility of extending the reach of the LC
to higher values of $m_{A^0,H^0,H^{\pm}} \gtrsim 0.5 \sqrt{s}$.  
Single heavy Higgs boson production has been studied in the context of
the MSSM or a general two Higgs doublet model (2HDM)
in a number of processes at LC, $\gamma\gamma$ collider and 
$e^-\gamma$ collider.  
At a LC, the following final states have been considered:\\
(I) Tree-level processes suppressed by 
$\cos^2(\beta - \alpha)$, where $\alpha$ is the mixing angle
that diagonalizes the CP-even neutral Higgs boson mass-squared matrix.
Such processes includes the associated $h^0A^0$ production 
$e^+e^-\rightarrow h^0A^0$\cite{FDHiggsprod},
Higgsstrahlung $e^+e^-\rightarrow ZH^0$\cite{FDHiggsprod}, 
$W$ and $Z$-boson fusion\cite{Orangebook},
$e^+e^- \to \nu \bar \nu H^0$;
$e^+e^- \to e^+ e^- H^0$.
In the decoupling limit, 
$\cos^2(\beta - \alpha) \propto m_Z^4/m_{A^0}^4$,
so these cross sections decrease rapidly as $m_{A^0}$ increases. 
\\
(II) Production in association with
pairs of third-generation fermions\cite{bbHAHpm,bbHA,bbHpm,KanemuraReview}: 
$e^+e^- \to b \bar b H^0$, $b \bar b A^0$,
$\tau^- \bar \nu H^+$, and $\bar t b H^+$.
The cross sections for the first
three of these processes are strongly enhanced at large $\tan\beta$, and 
the fourth is enhanced at both large and small 
$\tan\beta$.\\
(III) Single heavy MSSM Higgs production modes that are zero
at the tree level but arise at one loop:
$e^+e^- \to Z A^0$\cite{eeZA2HDM,eeZAMSSM}, 
$e^+e^- \to \gamma A^0$\cite{gammaA,eeZA2HDM,FGL}, 
$e^+e^- \to \nu \bar\nu A^0$\cite{FGL,nunuafull,arhrib} 
and $e^+e^- \to W^{\pm} 
H^{\mp}$\cite{eeWH2HDM,KanemuraeeHW,SHZhu,Breinee, HW, HW2}.

\begin{figure}
\resizebox{7.6 cm}{!}{
\includegraphics{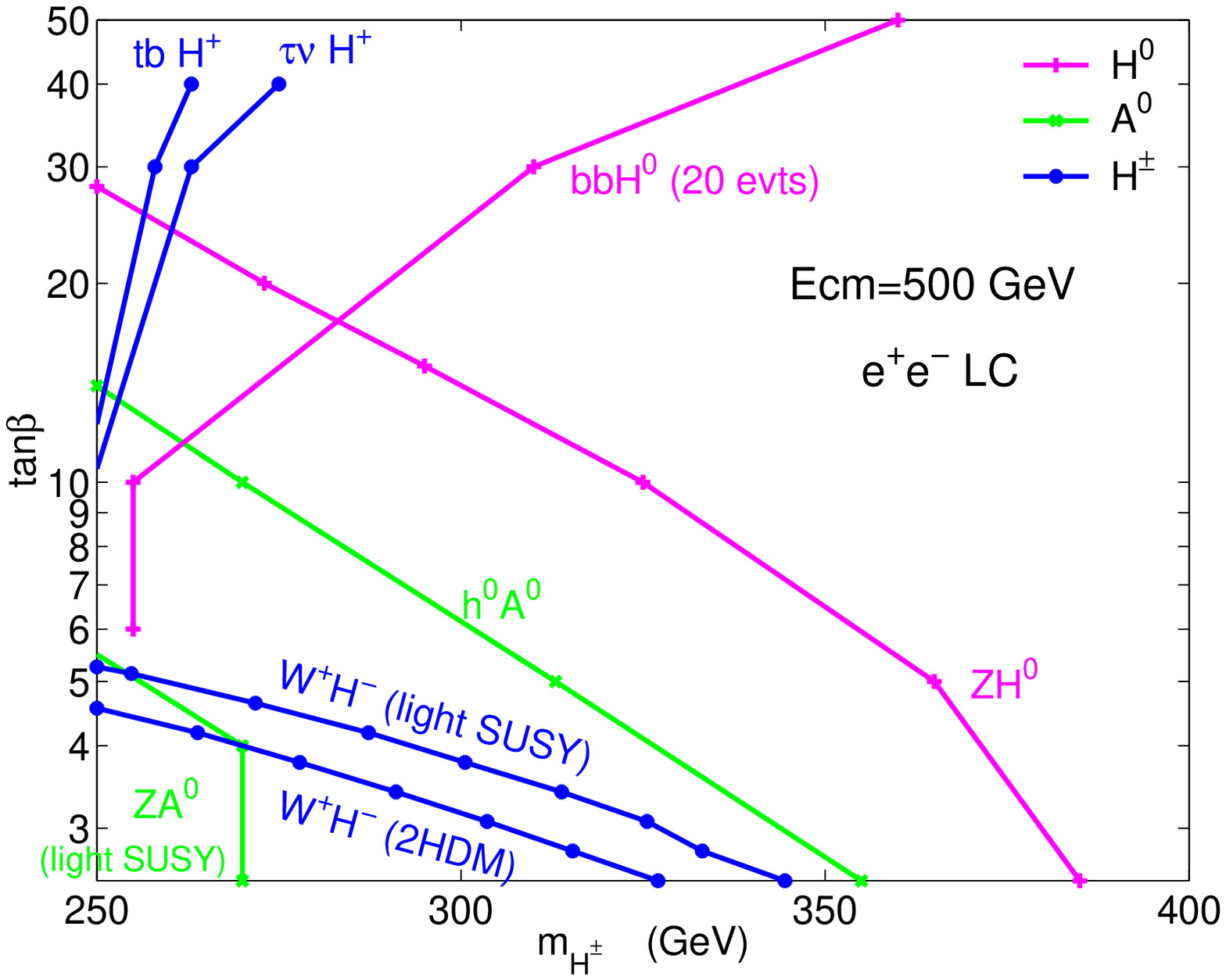}}
\resizebox{7.5 cm}{!}{
\includegraphics{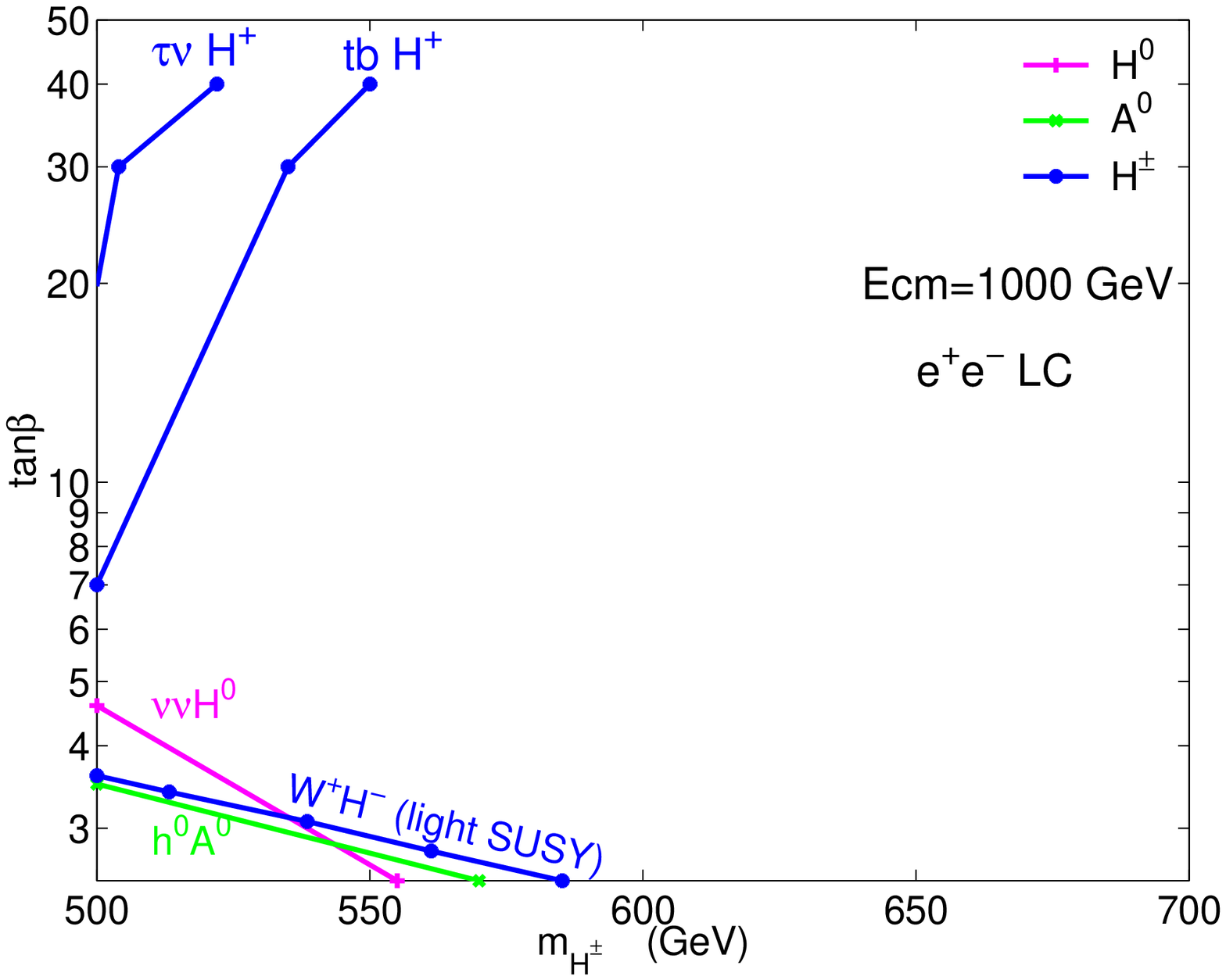}}
\caption{Ten-event contours for single heavy Higgs boson production in
unpolarized $e^+e^-$ collisions for $\sqrt{s}=500$ GeV (left), 
with 500 fb$^{-1}$ of integrated luminosity and $\sqrt{s}=1000$ GeV , 
with 1000 fb$^{-1}$ of integrated luminosity (right).
On the $x$-axis we plot the mass of the relevant heavy Higgs boson.
}
\label{fig:LC}
\end{figure}
Because detailed experimental studies of almost all of these processes
are unavailable, we choose an optimistic standard of detectability to be
10 heavy Higgs boson production events in the LC data sample.
We assume data samples of 500 fb$^{-1}$ at $\sqrt{s} = 500$ GeV
and 1000 fb$^{-1}$ at $\sqrt{s} = 1000$ GeV.  
For neutral Higgs boson production, this 10-event standard corresponds
to a cross section of 0.02 fb at $\sqrt{s} = 500$ GeV
(0.01 fb at $\sqrt{s} = 1000$ GeV).
For charged Higgs boson production, we add together the cross sections
for $H^+$ and $H^-$ production before applying the 10-event standard.
In what follows we assume that the $e^+$ and $e^-$ beams are unpolarized,
and adapt the cross sections presented in the literature accordingly.
We consider only $\tan\beta$ values above the LEP lower bound of 
2.4 \cite{LEP2} and heavy Higgs masses above $\sqrt{s}/2$.
The comparison of various production modes in the reach of $m_H-\tan\beta$
plane is given in Fig.~\ref{fig:LC}.

If the $e^+e^-$ LC is converted into a photon collider through Compton 
backscattering of intense laser beams, 
the neutral heavy Higgs bosons $H^0$ and $A^0$ can be singly produced in the 
$s$-channel through their loop-induced couplings to photon pairs.  
This process appears to be very promising for detecting 
$H^0$ and $A^0$ with masses above $\sqrt{s}/2$ 
and moderate $\tan\beta$ values between 2.5 and 10
\cite{gamgamAsner,gamgamMuhlleitner,gamgamGunionHaber}.
In particular, a recent realistic simulation of signal and 
backgrounds \cite{gamgamAsner} showed that
a 630 GeV $e^+e^-$ LC running in $\gamma\gamma$ mode
for three years would allow $H^0$, $A^0$ detection 
over a large fraction of the LHC wedge region (in which the heavy MSSM
Higgs bosons would not be discovered at the LHC)
for $m_{A^0}$ up to the photon-photon energy limit of 
$\sim 500$ GeV.  
At a 1000 GeV LC, the mass reach is likely to be above 
600 GeV \cite{gamgamMuhlleitner}.

The cross sections for production of $\tau^- \bar \nu H^+$ and 
$\bar t b H^+$ in $\gamma\gamma$ collisions \cite{CPYuan} are expected to be 
larger than the corresponding cross sections in $e^+e^-$ collisions
at large $\tan\beta$.
Production of $W^+H^-$ in $\gamma\gamma$ collisions also 
occurs at the one-loop level\cite{ggHW}, which is 
competitive with $e^+e^- \to W^+H^-$ in the MSSM for a TeV machine. 

Finally, if the LC is run in $e^- \gamma$ mode, the process
$e^- \gamma \to \nu H^-$ is possible.  
Unfortunately, with the typical expected $e^- \gamma$ 
luminosity of 100 fb$^{-1}$,
the cross section for this process in the 2HDM is too small to be of interest
\cite{Kanemuraegamma}.
This process could become promising in the MSSM if its cross section
is enhanced by the contributions of light superpartners, or if the
$e^- \gamma$ luminosity is increased.

Let us now focus on the two processes that we have studied: 
$e^+e^-\rightarrow W^{\pm}H^{\mp}$ and 
$e^+e^-\rightarrow \nu\bar\nu A^0$.  
The 2HDM contributions to $e^+e^-\rightarrow W^{\pm}H^{\mp}$
have been studied in \cite{eeWH2HDM}, and we calculated the additional
contributions from superparticles in MSSM.  Details of the analysis 
can be found in \cite{HW}.  Similar analysis has been done \cite{Breinee}
and reported by O.~Brein at this conference. 
There is no contribution at tree level due to the vanishing of 
$\gamma W^+ H^-$ and $Z W^+ H^-$ couplings and the leading contribution 
appears at one-loop.  
Unlike the case of the non-supersymmetric 2HDM, 
in which the top/bottom quark loops give by far the 
largest contribution to the cross section,
in the full MSSM the fermionic loops
involving charginos/neutralinos and the bosonic loops involving 
stops/sbottoms also give contributions of similar size.  
Although the stop/sbottom loops are enhanced by 
the large $H^{-}\tilde{t}_R\tilde{b}_L^*$ coupling
(which is proportional to the top quark Yukawa coupling), 
these diagrams are suppressed by higher powers of the superparticle 
masses than the fermionic loops.

\begin{figure*}
\resizebox{15.5cm}{!}{
\includegraphics{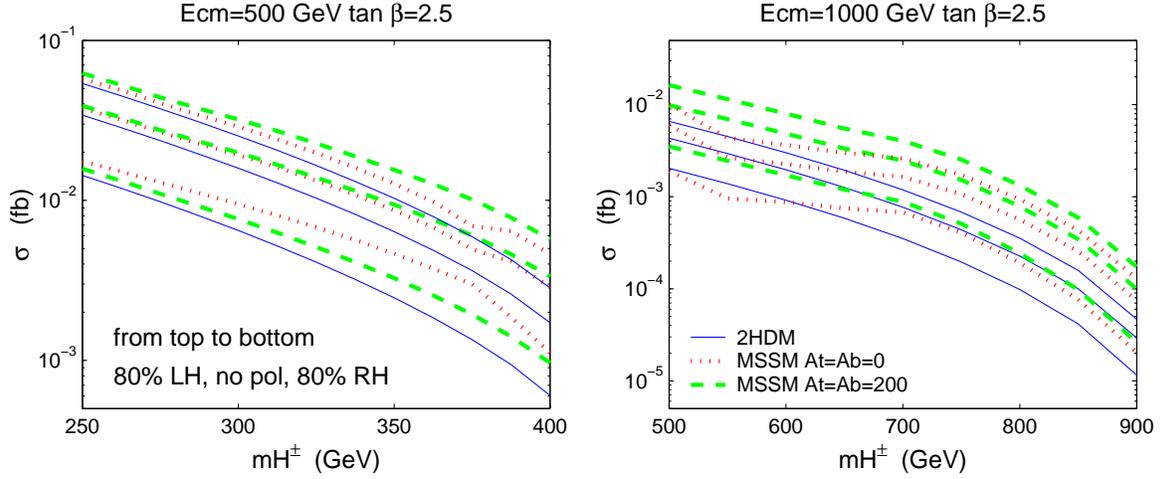}}
\caption{The $e^+e^-\rightarrow{W^+H^-}$ cross section as a function
of $m_{H^{\pm}}$ for $\tan\beta=2.5$, at 
$\sqrt{s} = 500$ GeV (left) and 1000 GeV (right).
The trilinear couplings are chosen as $A_t=A_b=0$ (dotted lines)
and 200 GeV (dashed lines).
The rest of the SUSY parameters are chosen to be
$M_{\rm SUSY}=200$ GeV, $2M_1=M_2=200$ GeV, and $\mu=500$ GeV.
The solid lines show the cross section in the non-SUSY 2HDM 
(with MSSM relations for the Higgs sector).}
\label{fig:mHpm2.5}
\end{figure*}
Figures~\ref{fig:mHpm2.5} show the dependence of the 
$e^+e^-\rightarrow{W^+H^-}$ cross section on the charged Higgs mass
for $\tan\beta = 2.5$, 
$\sqrt{s}=500$ GeV (left) and $\sqrt{s}=1000$ GeV (right).
Solid lines are the contributions from the non-SUSY
2HDM with the Higgs sector constrained by the MSSM mass and coupling 
relations.
The dotted (dashed) lines show the cross sections in the full MSSM,
including the contributions from all the superparticles,
for $A_t=A_b=0$ ($A_t=A_b=200$ GeV).
We compare the cross sections with
80\% left-handed $e^-$ polarization, no polarization, and 80\% right-handed
$e^-$ polarization, which are denoted in each plot by the same type of lines,  
from top to bottom.  
Left-handed $e^-$ polarization always gives a larger cross section.   
The additional SUSY contributions generally enhance the cross section.  
The cross sections decline as $m_{H^{\pm}}$ increases; however,
reasonable cross sections can be obtained for $m_{H^{\pm}} > \sqrt{s}/2$,
especially for small $\tan\beta$.

\begin{figure*}
\resizebox{15.5cm}{!}{
\includegraphics{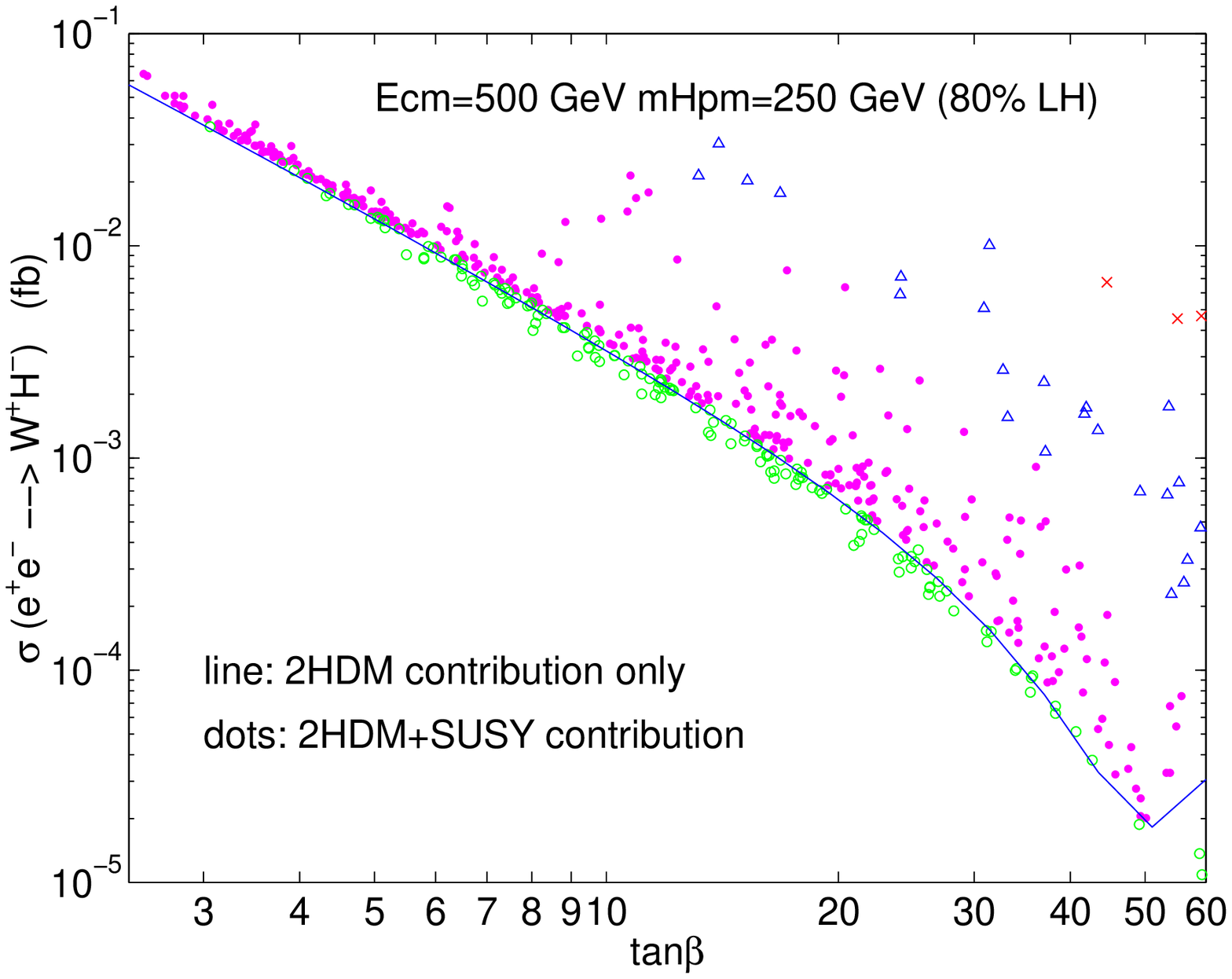}
\includegraphics{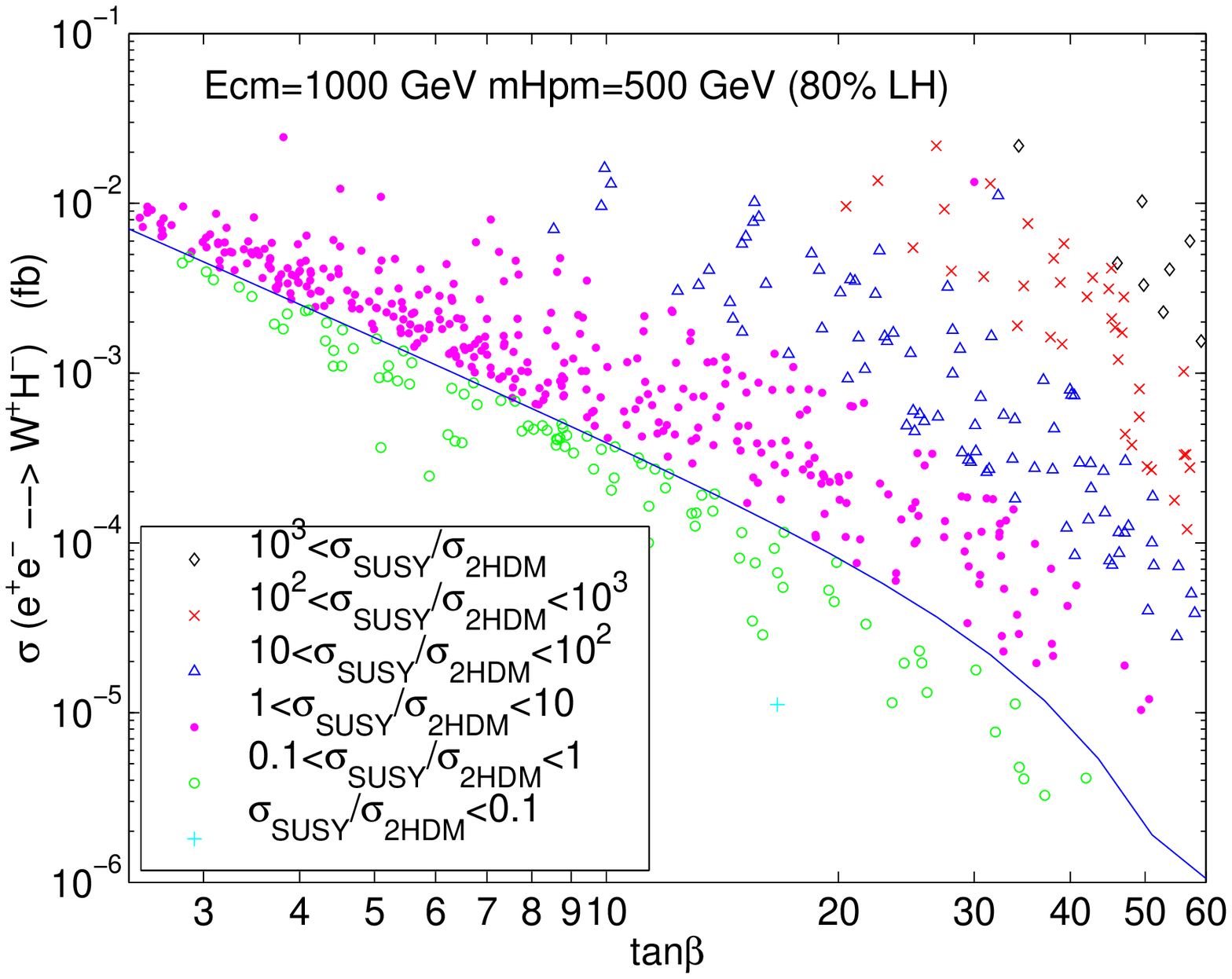}}
\caption{MSSM cross section as a function of $\tan\beta$ for
$\sqrt{s} = 500$ GeV and $m_{H^{\pm}} = 250$ GeV (left) 
and $\sqrt{s} = 1000$ GeV and $m_{H^{\pm}} = 500$ GeV (right).
The different symbols show the enhancement of the MSSM cross section
relative to the 2HDM.
The solid line shows the 2HDM cross section.
}
\label{fig:tanbeta}
\end{figure*}
In Fig.~\ref{fig:tanbeta} we compare the cross section in the MSSM with that
in the 2HDM as a function of $\tan\beta$, 
for $\sqrt{s} = 500$ and 1000 GeV.  We fix $m_{H^{\pm}}$ to a single value, 
$m_{H^{\pm}} = \sqrt{s}/2$, in order to more clearly illustrate the 
effect of the MSSM contributions.  The rest of the MSSM parameters are 
chosen randomly within 1 TeV, taking into account the 
experimental lower limit on the superparticle searches.  
Details of the analysis can be found at \cite{HW2}.
While the 2HDM cross section falls rapidly with increasing $\tan\beta$,
the MSSM contributions depend much more weakly on $\tan\beta$,
especially at $\sqrt{s} = 1000$ GeV, where the maximum cross section is
almost independent of $\tan\beta$.  This implies that the
largest relative cross section enhancements due to MSSM contributions 
occur at large $\tan\beta$, as shown by the different symbols in 
Fig.~\ref{fig:tanbeta}.  At a 500 GeV machine, an enhancement of more than
a factor of 10 can occur for $\tan\beta > 10$, while a factor of
100 can occur for $\tan\beta > 40$.  At a 1000 GeV machine,
the enhancements can be even larger.
At low $\tan\beta$, cross section enhancements of roughly 50\% are typical.

\begin{figure*}
\resizebox{15.5cm}{!}{
\includegraphics{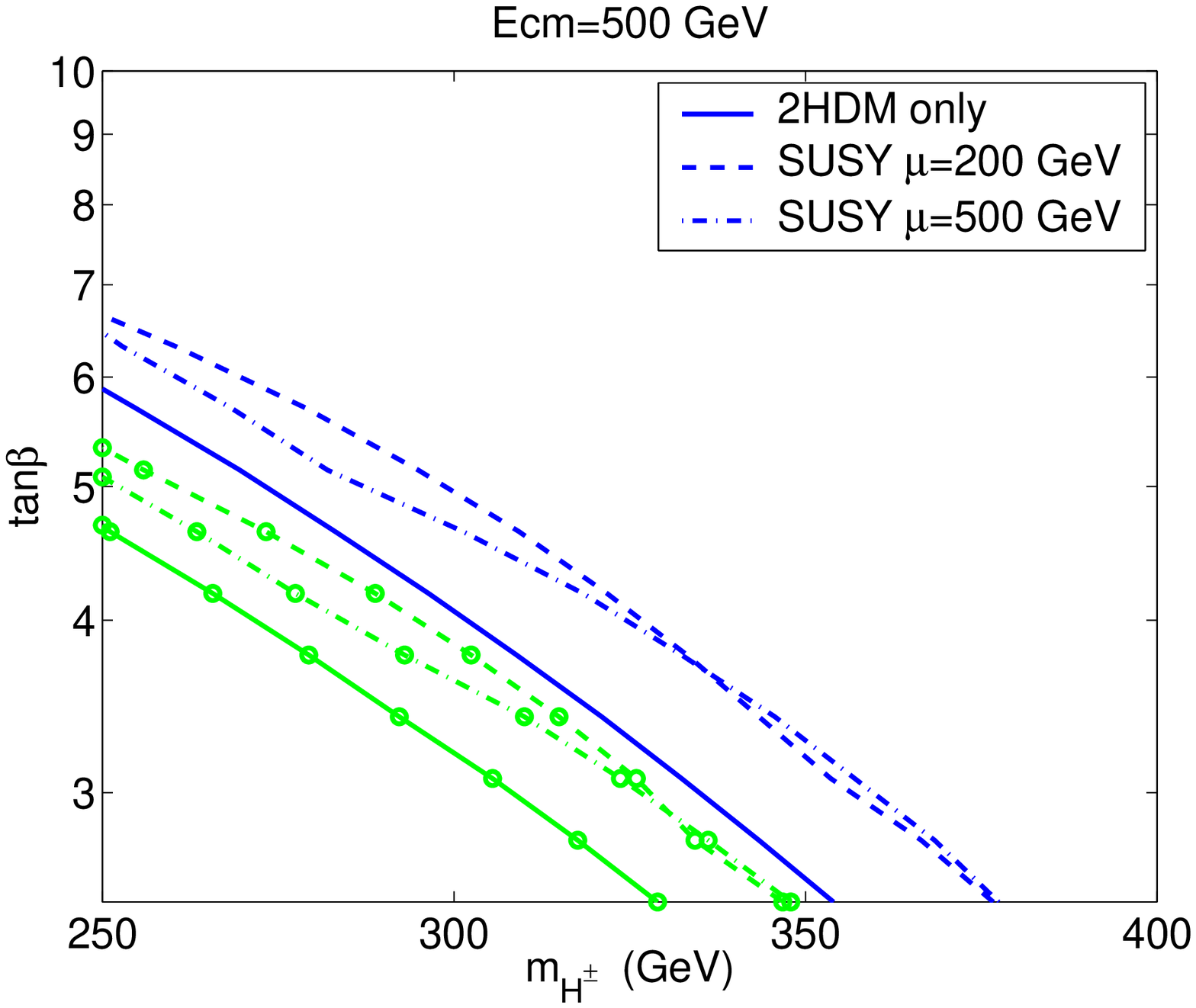}
\includegraphics{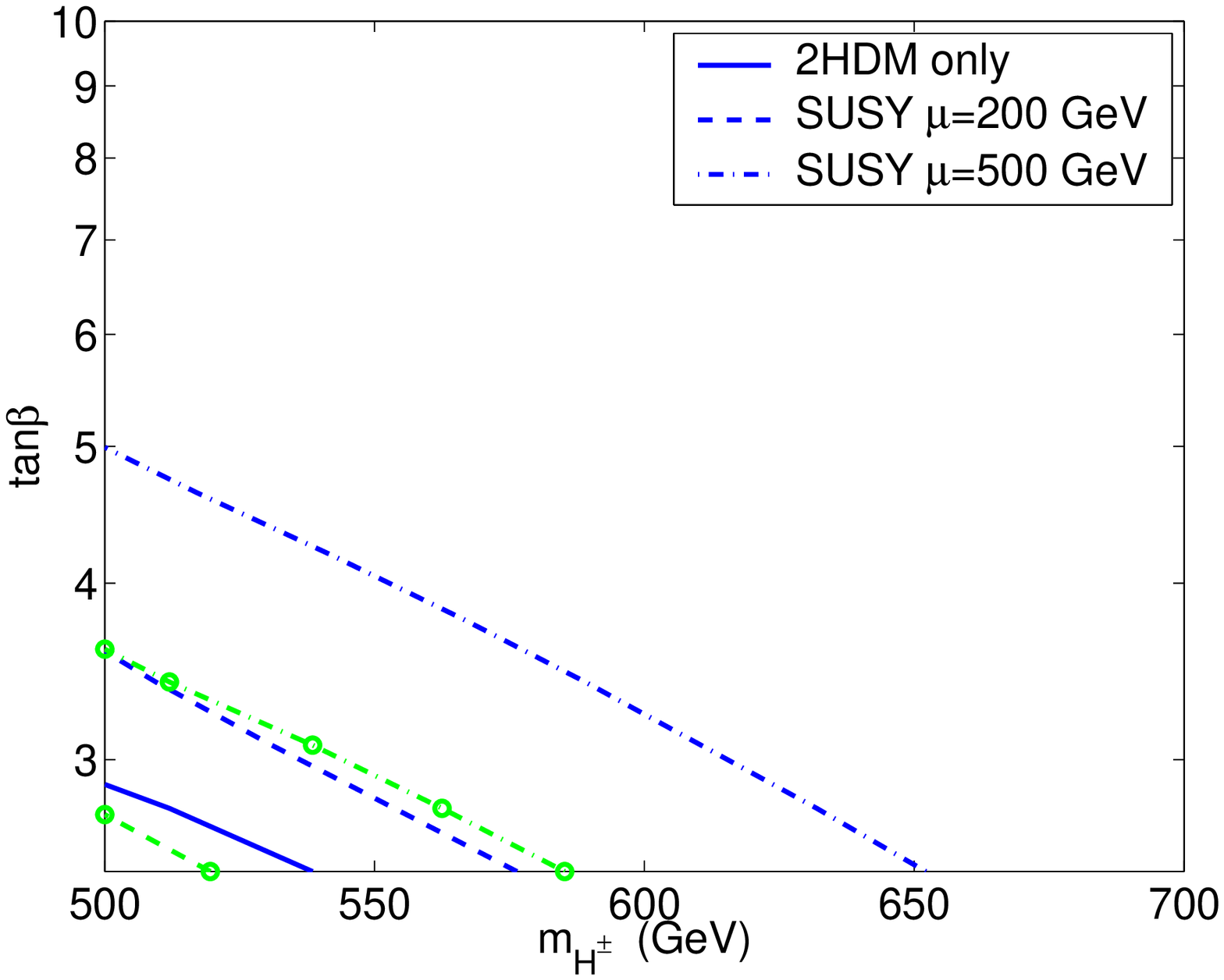}}
\caption{Ten-event contours for $e^+e^- \to W^{\pm}H^{\mp}$
for $\sqrt{s} = 500$ GeV with 500 fb$^{-1}$ (left)
and $\sqrt{s} = 1000$ GeV with 1000 fb$^{-1}$ (right).
The SUSY parameters are chosen to be $M_{\rm SUSY}^{tb} = 1000$ GeV for the
top and bottom squarks, $M_{\rm SUSY} = 200$ GeV for the rest of the squarks
and the sleptons, $2M_1 = M_2 = 200$ GeV, $M_{\tilde g} = 800$ GeV,
$A_t = A_b = 2 M_{\rm SUSY}^{tb}$,
and $\mu = 200$ GeV (dashed lines) and 500 GeV (dot-dashed lines).
Solid lines show the 2HDM result. 
Light (green) lines show the unpolarized cross section and dark (blue) lines
show the cross section with an 80\% left-polarized $e^-$ beam.
}
\label{fig:contours}
\end{figure*}

The maximum 10-event reaches in $m_{H^{\pm}}$ and $\tan\beta$ is
shown in Fig.~\ref{fig:contours}.
In all cases considered, the MSSM contributions increase the 10-event 
reach over that in the 2HDM:
the reach in $m_{H^{\pm}}$ is increased by about 20 GeV at 
$\sqrt{s} = 500$ GeV, and by about 40 GeV or more at $\sqrt{s} = 1000$ GeV.
At $\sqrt{s} = 1000$ GeV, increasing the $\mu$ parameter from 200 to 
500 GeV with other SUSY parameters held fixed increases the reach by 
about 70 GeV in $m_{H^{\pm}}$, and by about 1 unit in $\tan\beta$.  
At $\sqrt{s} = 500$ GeV, the $\mu$ dependence is much weaker.
For either value of $\mu$,
using an 80\% left-polarized $e^-$ beam increases the reach compared to
using unpolarized beams by about 30 GeV in $m_{H^{\pm}}$ at 
$\sqrt{s} = 500$ GeV and by twice that at $\sqrt{s} = 1000$ GeV; at either
center-of-mass energy the reach in $\tan\beta$ increases by about 1 unit.

Comparing to the other single charged Higgs production modes at 
LC collider, we see that while 
$e^+e^-\rightarrow \tau\bar\nu H^+$ and $e^+e^-\rightarrow \bar{t} b H^+$
is promising at large $\tan\beta$, 
$W^+H^-$ production is the only channel
in $e^+e^-$ collisions
analyzed to date that yields $\geq 10$ events containing charged Higgs bosons
at low $\tan\beta$ values for $m_{H^{\pm}}\geq \sqrt{s}/2$.

\begin{figure*}
\resizebox{15.5cm}{!}{
\rotatebox{270}{\includegraphics{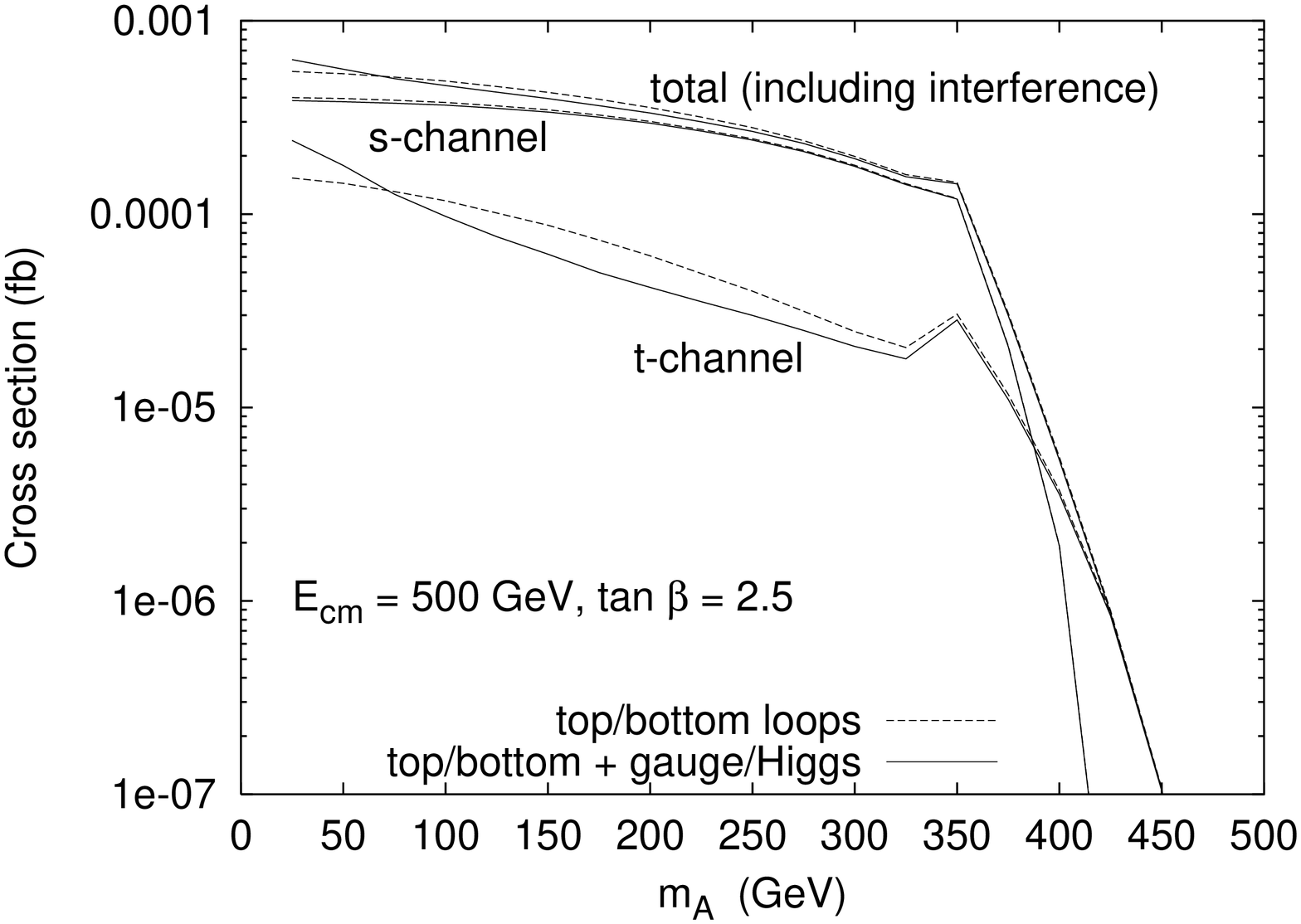}}
\rotatebox{270}{\includegraphics{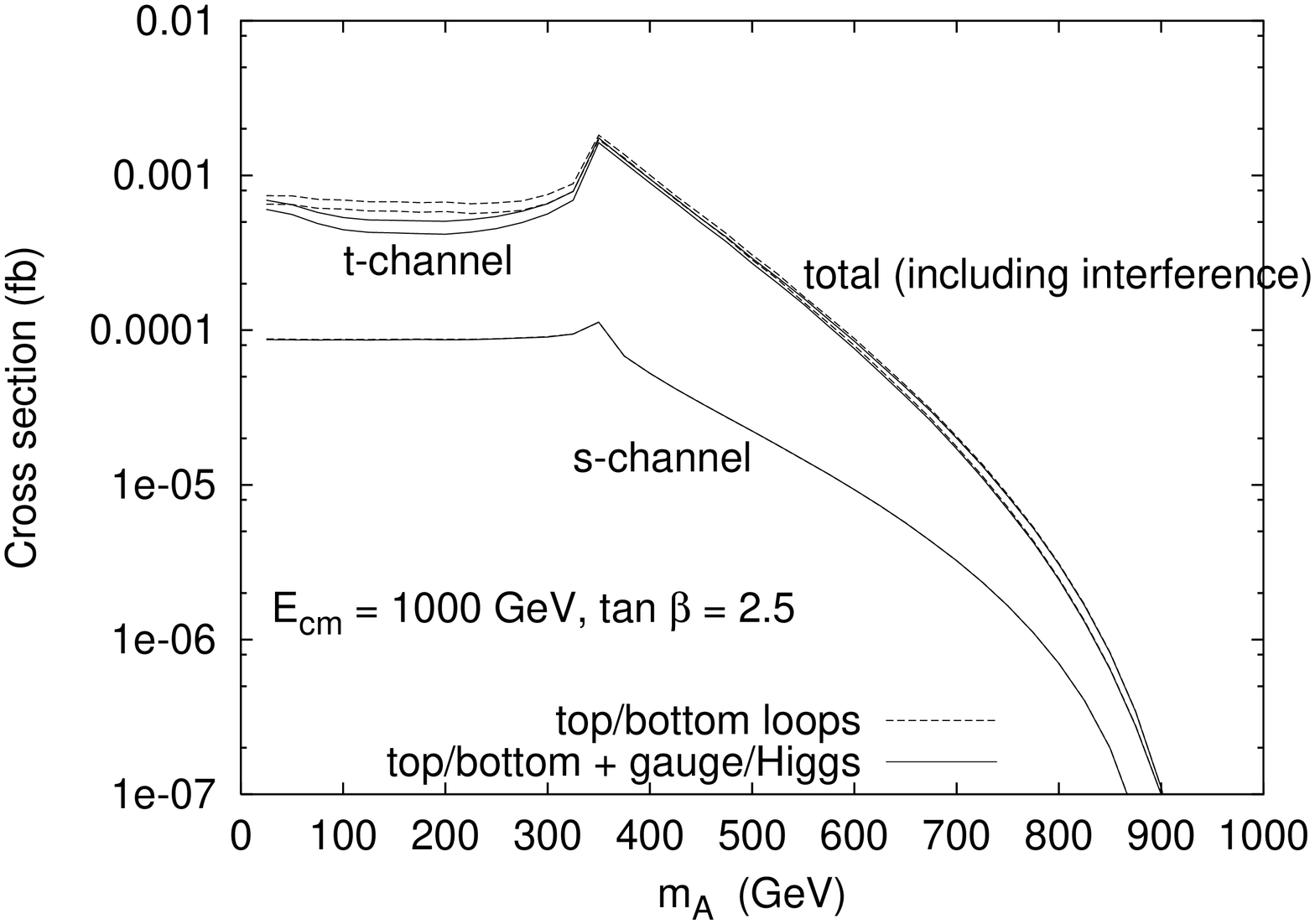}}}
\caption{Cross section of $e^+e^- \to \nu \bar\nu A^0$
for $\tan\beta=2.5$, $\sqrt{s} = 500$ GeV (left)
and $\sqrt{s} = 1000$ GeV (right) with $s$-channel, $t$-channel and 
$s+t$ contributions, respectively.  Solid lines show the total contributions 
from top/bottom quarks, Higgs bosons and gauge bosons, while the dashed
lines show the top/bottom quark contribution only.  
Non-supersymmetric 2HDM contributions
are included and MSSM mass relation for the Higgs sector is imposed.  
}
\label{fig:nunuA}
\end{figure*}
For the single CP-odd Higgs production mode 
$e^+e^-\rightarrow \nu\bar\nu A^0$, the leading 
contribution from top and bottom quark correction to the 
$WWA^0$ vertex has been reported in \cite{FGL}.
Details of the full 2HDM contributions can be found in \cite{nunuafull}.  
Part of the SUSY contributions 
including CP violation phases has been done in \cite{arhrib} and 
reported by A.~Arhrib at this conference.  
In addition to the  $t$-channel $W$ fusion contributions, 
there are diagrams of $s$-channel Z boson exchange.
For on-shell intermediate $Z$, such process is 
the same as $e^+e^-\to Z A^0$ with $Z$ decaying into neutrinos, 
which has been calculated for 2HDM and full
MSSM in \cite{eeZA2HDM, eeZAMSSM} and the 
production cross section is found to be small. 
Moreover, $Z$ resonance portion can be removed experimentally 
by the recoil reconstruction.  
However, for off-shell intermediate $Z^*$, 
such processes have the same
$\nu\bar\nu A^0$ final state, which interfere with the $t$-channel diagrams
and have to be included in our calculation.
Fig.~\ref{fig:nunuA} show the cross section for 2HDM contributions, 
for 500 GeV and TeV LC.  
The cross section is too small to be observed for 
$m_{A^0} > \sqrt{s}/2$ \cite{FGL}.

In summary, we can see that the single heavy Higgs production could 
be important at LC when the heavy Higgs mass is larger than
half of the center of mass energy.  The production 
modes $e^+e^-\rightarrow ZH^0, h^0A^0, W^{\pm}H^{\mp}$ and $b\bar{b}H^0$
at 500 GeV machine and $e^+e^-\rightarrow W^{\pm}H^{\mp}$, $h^0A^0$, 
$\nu\bar\nu H^0$ and $t\bar{b}H^-$ at TeV 
machine has reasonable 10-event reach in $m_{H^{\pm}}-\tan\beta$ plane. 
At low $\tan\beta$, SUSY contributions from light superparticles and 
using left-polarized electron beam could enhance the 
cross section for $e^+e^-\rightarrow W^{\pm}H^{\mp}$.  
While for $e^+e^-\rightarrow \nu\bar\nu A^0$, the non-supersymmetric 2HDM 
contribution seems to be too small to be interesting.  
A few additional processes that have not yet been computed may be 
promising for single heavy Higgs boson production at an $e^+e^-$ collider,
for example, the weak boson fusion process $e^+e^- \to \bar \nu e^- H^+$
at $\sqrt{s} \sim 1000$ GeV.
Also, to have a realistic analysis of the reach for 
various single heavy Higgs production modes, the background study 
is important and necessary.

{\sl Acknowledgment:} 
The author is grateful to Heather Logan, Jack Gunion and Tom Farris 
for collaborations on works presented in this talk.  We would also like 
to thank Oliver Brein and Abdesslam Arhrib for comparison of the results. 
The work of S.S is supported by 
the DOE under grant DE-FG03-92-ER-40701 and 
by the John A. McCone Fellowship.


\begin{thebibliography}{99}



\bibitem{HiggsatRunII}
M.~Carena {\it et al.},
Report of the Tevatron Higgs working group,
arXiv:hep-ph/0010338.

\bibitem{GuchaitHpm}
M.~Guchait and S.~Moretti,
JHEP {\bf 0201}, 001 (2002).


\bibitem{AtlasTDR}
K.~Lassila-Perini, ETH Dissertation thesis No. 12961 (1998);
ATLAS collaboration Technical Design Report, available from
\verb+http://atlasinfo.cern.ch/Atlas/+
\verb+GROUPS/PHYSICS/TDR/access.html+.

\bibitem{CMS}
D.~Denegri {\it et al.},
arXiv:hep-ph/0112045.

\bibitem{AtlasH+}
K.~A.~Assamagan, Y.~Coadou and A.~Deandrea,
arXiv:hep-ph/0203121;
K.~A.~Assamagan and Y.~Coadou,
Acta Phys.\ Polon.\ B {\bf 33}, 707 (2002).


\bibitem{LEP2}
LEP Higgs Working Group Collaboration,
arXiv:hep-ex/0107030.


\bibitem{FDHiggsprod}
S.~Heinemeyer, W.~Hollik, J.~Rosiek and G.~Weiglein,
Eur.\ Phys.\ J.\ C {\bf 19}, 535 (2001);
see \verb+www.feynhiggs.de+.

\bibitem{Orangebook}
T.~Abe {\it et al.}  [American Linear Collider Working Group Collaboration],
{\it Linear collider physics resource book for Snowmass 2001,
Part 2: Higgs and  supersymmetry studies,}
hep-ex/0106056.


\bibitem{bbHAHpm}
A.~Gutierrez-Rodriguez, M.~A.~Hernandez-Ruiz and O.~A.~Sampayo,
arXiv:hep-ph/0110289.

\bibitem{bbHA}
U.~Cotti, A.~Gutierrez-Rodriguez, A.~Rosado and O.~A.~Sampayo,
Phys.\ Rev.\ D {\bf 59}, 095011 (1999).


\bibitem{bbHpm}
A.~Gutierrez-Rodriguez and O.~A.~Sampayo,
Phys.\ Rev.\ D {\bf 62}, 055004 (2000).

\bibitem{KanemuraReview}
S.~Kanemura, S.~Moretti and K.~Odagiri,
JHEP {\bf 0102}, 011 (2001).

\bibitem{eeZA2HDM}
A.~G.~Akeroyd, A.~Arhrib and M.~Capdequi Peyran\`ere,
Mod.\ Phys.\ Lett.\ A {\bf 14}, 2093 (1999)
[Erratum-ibid.\ A {\bf 17}, 373 (2002)].

\bibitem{eeZAMSSM}
A.~G.~Akeroyd, A.~Arhrib and M.~Capdequi Peyran\`ere,
Phys.\ Rev.\ D {\bf 64}, 075007 (2001)
[Erratum-ibid.\ D {\bf 65}, 099903 (2002)].



\bibitem{gammaA}
A.~Djouadi, V.~Driesen, W.~Hollik and J.~Rosiek,
Nucl.\ Phys.\ B {\bf 491}, 68 (1997).


\bibitem{FGL}T.~Farris, J.~F.~Gunion and H.~E.~Logan, Contributions
to the snowmass 2001 Workshop on ``The Future of Particle Physics'', Snowmass,
CO, USA, July 2001 [arXiv:hep-ph/0202087].

\bibitem{nunuafull} T.~Farris, J.~F.~Gunion, H.~E.~Logan and S.~Su, 
in preparation.

\bibitem{arhrib}A.~Arhrib, talk given at ``SUSY02, 10th international 
conference
on Supersymmetry and unification of fundamental interactions'', June 17-23, 
2002, DESY Hamburg.


\bibitem{eeWH2HDM}
A.~Arhrib, M.~Capdequi Peyran\`ere, W.~Hollik and G.~Moultaka,
Nucl.\ Phys.\ B {\bf 581}, 34 (2000).

\bibitem{KanemuraeeHW}
S.~Kanemura,
Eur.\ Phys.\ J.\ C {\bf 17}, 473 (2000).

\bibitem{SHZhu}
S.~H.~Zhu,
arXiv:hep-ph/9901221.

\bibitem{Breinee}
O.~Brein, W.~Hollik and T.~Hahn, in preparation;
O.~Brein,
arXiv:hep-ph/0209124.

\bibitem{HW}H.~E.~Logan and S.~Su, Phys.\ Rev.\ D {\bf 66}, 035001 (2002).

\bibitem{HW2}H.~E.~Logan and S.~Su, arXiv:hep-ph/0206135.


\bibitem{gamgamAsner}
D.~M.~Asner, J.~B.~Gronberg and J.~F.~Gunion,
arXiv:hep-ph/0110320.

\bibitem{gamgamMuhlleitner}
M.~M.~M\"uhlleitner, M.~Kramer, M.~Spira and P.~M.~Zerwas,
Phys.\ Lett.\ B {\bf 508}, 311 (2001).

\bibitem{gamgamGunionHaber}
J.~F.~Gunion and H.~E.~Haber,
Phys.\ Rev.\ D {\bf 48}, 5109 (1993).

\bibitem{CPYuan}
H.~J.~He, S.~Kanemura and C.~P.~Yuan,
arXiv:hep-ph/0203090.

\bibitem{ggHW}
F.~Zhou, W.~G.~Ma, Y.~Jiang, X.~Q.~Li and L.~H.~Wan,
Phys.\ Rev.\ D {\bf 64}, 055005 (2001).

\bibitem{Kanemuraegamma}
S.~Kanemura and K.~Odagiri,
arXiv:hep-ph/0104179.




\end{thebibliography}
\end{document}